\documentclass[journal,comsoc]{IEEEtran}
\usepackage[T1]{fontenc} % Use 8-bit encoding that has 256 glyphs
\usepackage[utf8]{inputenc}
\usepackage[english]{babel} % English language/hyphenation
\usepackage{amsmath,amsfonts, amssymb} % Math packages
\usepackage{mathtools}
\usepackage{acronym}
\usepackage{hyperref}
\usepackage{longtable}
\usepackage{textcomp}
\usepackage{multirow}
\usepackage{makecell}
\usepackage{subfigure}
\usepackage{graphicx}
\usepackage{acronym}
\usepackage{color}
\usepackage{epsfig}
\usepackage{url}
\usepackage{balance}
\usepackage{float}
\usepackage{array}
\usepackage{comment}
\usepackage{pifont}
\usepackage{verbatim}
\usepackage{pifont}
\usepackage{tabularx}
\usepackage{hyperref}
\usepackage{comment}

\newcolumntype{P}[1]{>{\centering\arraybackslash}p{#1}}

\acrodef{API}{Application Program Interface}
\acrodef{BLE}{Bluetooth Low Energy}
\acrodef{CA}{Certification Authority}
\acrodef{SDR}{Software Defined Radio}
\acrodef{DoS}{Denial of Service}
\acrodef{DH}{Diffie Hellman}
\acrodef{ECC}{Elliptic Curve Cryptography}
\acrodef{ECDH}{Elliptic Curve Diffie Hellman}
\acrodef{GNSS}{Global Navigation Satellite System}
\acrodef{GPS}{Global Positioning System}
\acrodef{IoT}{Internet of Things}
\acrodef{KDF}{Key Derivation Function}
\acrodef{HKDF}{HMAC Key Derivation Function}
\acrodef{MITM}{Man-in-the-Middle}
\acrodef{PKC}{Public Key Cryptography}
\acrodef{PKI}{Public Key Infrastructure}
\acrodef{RF}{Radio Frequency}
\acrodef{RFID}{Radio Frequency Identification}
\acrodef{RSS}{Received Signal Strength}
\acrodef{SNR}{Signal-to-Noise-Ratio}
\acrodef{TLS}{Transport Layer Security}
\acrodef{PEPP-PT}{Pan-European Privacy-Preserving Proximity Tracing}
\acrodef{TCN}{Temporary Contact Numbers}
\acrodef{UHF}{Ultra High Frequency}
\acrodef{TDMA}{Time Division Multiple Access}
\acrodef{GFSK}{Gaussian Frequency Shift Keying}
\acrodef{DPSK}{Differential Phase Shift Keying}
\acrodef{FHSS}{Frequency-Hopping Spread Spectrum}
\acrodef{PACT}{Private Automated Contact Tracing}
\acrodef{RSSI}{Received Signal Strength Indicator}
\acrodef{DP-3T}{Decentralized Privacy-Preserving Proximity Tracing}
\acrodef{GDPR}{General Data Protection Regulation}

\graphicspath{{figs/}}

\title{IoTrace: A Flexible, Efficient, and Privacy-Preserving IoT-enabled Architecture for Contact Tracing}
\author{
    \IEEEauthorblockN{Pietro Tedeschi, Spiridon Bakiras, and Roberto Di Pietro}\\
    \IEEEauthorblockA{
    Division of Information and Computing Technology \protect\\
    College of Science and Engineering\\
    Hamad Bin Khalifa University,  Doha - Qatar
    \\Email: \{ptedeschi, sbakiras, rdipietro\}@hbku.edu.qa}}

\begin{document}

\maketitle

% \excludecomment{abstract}
% \excludecomment{figure}
% \let\endfigure\relax
% \excludecomment{table*}
% \let\endtable\relax

\begin{abstract}
Contact tracing promises to help fight the spread of \textsc{Covid-19} via an early detection of possible contagion events. \textcolor{black}{To this end, most existing solutions share the following architecture: smartphones continuously broadcast random beacons that are intercepted by nearby devices and stored into their local contact logs.} In this paper, we propose an IoT-enabled architecture for contact tracing that relaxes the smartphone-centric assumption, and provide a solution that enjoys the following features: (i) it reduces the overhead on the end-user to the bare minimum---the mobile device only broadcasts its beacons; (ii) it provides the user with a degree of privacy not achieved by competing solutions---even in the most privacy adverse scenario, the solution provides $k$-anonymity; and, (iii) it is flexible: the same architecture can be configured to support several models---ranging from the fully decentralized to the fully centralized ones\textcolor{black}{---and the system parameters can be tuned to support the tracing of several social interaction models.} What is more, our proposal can also be adopted to tackle future human-proximity transmissible  diseases.
 %can handle new viruses with entirely different characteristics. 
 \textcolor{black}{Finally, we also highlight open issues and discuss a number of future research directions at the intersection of IoT and contact tracing.} %We believe that the novelty of the proposal, as well as its striking properties and flexibility, has the potential to pave the way for further research.
%start a novel research thread.

%Contact tracing via mobile applications has recently gained momentum, as one of the most appealing new attractive solutions in the proximity testing research domain. Although several social distancing apps have been developed to face the \textsc{Covid-19} pandemic, they could be subject to potential privacy breaches and security attacks. To this end, the contribution of this work is twofold. First, we provide an overview and a comparison of the emerging proximity testing protocols with an in-depth analysis of the security and privacy features. The second contribution is the design of a novel contact tracing architecture leveraging the edge computing paradigm guaranteeing the right balance between privacy and public safety. Further, we show that our approach can be extended towards (i) a fully distributed version and (ii) privacy-oriented architecture, thus providing the security and privacy research challenges and a few future directions.
\end{abstract}

\section{Introduction}
\label{sec:intro}
%\cite{cho2020contact}
% Figure with overall communication architecture, moved here for template reasons
One thing is clear about the \textsc{Covid-19} pandemic declared in March 2020: despite the release of few vaccines, the fight against the virus could still last for years---due to the required global mass production, untested efficacy at scale, expected delays in distribution, and the very same virus  polymorphic capabilities. Indeed, the initial battles gained against the virus 
%such as the increase in the number of cases in the USA (passed from a peak of $38,509$ in April to a high of $89,048$ in October)~\cite{who20}, 
\textcolor{black}{have been later lost, with the ``second wave'' ravaging the world as of November 2020 ~\cite{jhu20}.}
%a record number of $99,000+$ new cases per day in the USA only in the last days of October  2020~\cite{nytcovid20}.

The initial, dramatic spread of \textsc{Covid-19} prompted individual states and international organizations to implement drastic measures to ``flatten the curve'' of the pandemic~\textcolor{black}{~\cite{Ting2020, Cecilia2020}}. {\em Digital contact tracing} is one of the most promising technological solutions, and its premise is quite intuitive: leverage the user's smartphone to keep track of other users nearby (called \textit{contacts})~\cite{Garg2020}. Then, if a contact has a positive diagnosis for the coronavirus, the user is notified to take precautionary measures, such as testing or self-quarantine. The most prominent approach to contact tracing is to have each mobile device broadcast pseudo-random beacons via its \ac{BLE} interface.  These beacons are then received and recorded by other users within the \ac{BLE} transmission range. %~\cite{Nieminen2014}. 
\textcolor{black}{Alternatively, solutions like Israel's Hamagen~\cite{Hamagen} adopt the \ac{GNSS} for localization and proximity tracing.}

A watershed difference in contact tracing applications lies in the {\em reconciliation} process, i.e., the identification of ``infected'' beacons inside a user's contact list that signal possible contagion events. To one extreme, {\em centralized solutions} require all users to share their beacons and/or contact lists with the health authorities, who perform the reconciliation process and notify the exposed users. To the other extreme, {\em decentralized solutions} \textcolor{black}{do not collect any information from the mobile devices. Instead, when a user is diagnosed as positive, the app releases the user's beacons to the authorities, which are then distributed to all the other users in the system. As such, the app is responsible for the reconciliation process, by matching the released beacons against the stored contact logs.} 

% What the above described seems quite linear and reasonable.
\textcolor{black}{This generic contact tracing framework raises some concerns about the {\em usability} of the solution, and opens up %as well as 
the Pandora's box of {\em privacy} and {\em security} issues.} The cited dimensions, other than being critical on their own, could also thwart the widespread adoption of contact tracing, making it irrelevant in fighting the pandemic. \textcolor{black}{Indeed, Oxford researchers have calculated that, to be effective, contact tracing apps must be actively used by at least $60\%$ of the  population~\cite{Tang20}.}
%RDP there are no update on the adoption % of contact tracing for having it being useful?
%Pietro: No, we discussed this aspect in the response letter with a lot of details about it (e.g. see the Ofxord research). I also put a lot of details that in the paper have been removed.
\textcolor{black}{%While the exact percentage threshold might be up for debate, the fact remains that
To reach the above goal, a more usable and privacy-preserving solution would have the potential to attract more active users, thus increasing the effectiveness of contact tracing.}

\textcolor{black}{In terms of usability, the main challenges are related to the energy efficiency and computational cost 
of the contact tracing app. For example, one of the common criticisms against existing applications is the diminished smartphone battery life. While some energy is consumed on the periodic transmission of the device's beacon, the main factor behind battery drain  %the short battery life 
is the continuous scanning of the Bluetooth channel for beacons transmitted by the surrounding devices.} 
%Furthermore, the computational cost for actively maintaining detailed contact logs and performing the exposure notification functions could be overwhelming for low-cost mobile devices, thus interfering with regular usage.}
%On the other hand, issues like UX interface and design are not very important, because such apps are designed to simply run in the background.}
%RDP the above paragraph is commented since it does not add much to the point brought forward
%In fact, all available contact tracing applications have been designed to run on users' smartphones. Signaling one's presence is realized using pseudo-random beacons spread via Bluetooth while keeping track of the experienced contacts is realized via receiving and storing someone else's spread beacons. This implies an energetic cost that, though the \ac{BLE} protocol is used, cannot be neglected. A further cost is the required storage. As a matter of fact, while the beacons to be sent could be generated by a seed---hence having a seed enables a user to regenerate the sequence of spread beacons---the received beacons need to be individually stored. Between the two constraints, given the storage capability of smartphones and the actual size of a beacon ($64$-$128$ bits), the most relevant issue is the one related to energy consumption.

When it comes to privacy and security, the scientific community has started debating the issue from the very beginning~\cite{baumgrtner2020mind}. The most recurrent threats are user de-identification and user tracking. \textcolor{black}{In particular, an eavesdropper can identify a user as positive to the disease, by cross-referencing the ``infected'' beacons published by the authorities with the beacons acquired via eavesdropping. The same data may also allow an adversary to track the locations that a positively diagnosed individual has visited. This is a clear violation of the \ac{GDPR} laws in the EU and, in any case, a serious threat that could hinder the adoption of the contact tracing application.}

\textbf{Contributions.}
Motivated by the above observations, we introduce IoTrace, a contact tracing solution that relies on a distributed, flexible, and lightweight IoT-based infrastructure. IoTrace imposes minimal overhead on the user's smartphone, while providing strong privacy guarantees not available in competing proposals. Specifically, IoTrace relaxes the requirement for smartphones to receive the beacons issued by other devices in their proximity. This translates into considerable savings in energy consumption and computational/storage costs. Further advantages are that the IoT infrastructure is fully distributed, heterogeneous, and pervasive. Distribution and heterogeneity do help security \cite{Lu2019}, while pervasiveness would assure efficient and accurate contact tracing. The reconciliation mechanism is fully tunable and could range from a completely decentralized solution to a centralized one.
%RDP wording 
Finally, it is worth noting that the proposed solution would work for any type of \emph{human2human} transmissible disease---just tailoring the application parameters, such as the time of exposure needed to trigger the precautionary measures.

\section{Related Work}
\label{sec:related}
Several contact tracing applications have been developed in the last few months. 
%For a recent, though partial list of those ones, one can refer to \cite{}. 
In the following paragraphs, we provide a brief introduction of the state-of-the-art approaches and also present a quantitative comparison in terms of user privacy and performance.

\begin{table*}[htbp]
\caption{\color{black}Comparison of state-of-the-art representative solutions.
$n$: contact list size, $l$: number of stored locations, $f$: TX frequency, None: $-$, Low: $\star$, Medium: $\star\star$, High: $\star\star\star$. For IoTrace, some metrics include two ratings that correspond to the Basic: $\blacktriangle$ and Privacy-Enhanced: $\blacksquare$ versions.
}
\centering
    \color{black}{
    \begin{tabular}{|c||c|c|c|c|c|c|c|}
    \hline
        \textbf{Features} & BlueTrace~\cite{bluetrace} & DP-3T~\cite{troncoso2020} & Apple/Google~\cite{applegoogle} & Hamagen~\cite{Hamagen} & PEPP-PT~\cite{pepppt} & \multicolumn{2}{c|}{IoTrace} \\ \hline\hline
        \emph{Wireless Technology} & Bluetooth & Bluetooth & Bluetooth & GPS & Bluetooth & \multicolumn{2}{c|}{Bluetooth} \\ \hline
        \emph{Open-Source} & Yes & Yes & Yes & Yes & Yes & \multicolumn{2}{c|}{Yes} \\ \hline
        \emph{Architecture (C/D/H)} & H  & D  & D  & D  & H & C $\blacktriangle$ & D $\blacksquare$  \\ \hline
        \emph{RF Energy Consumption (mJ/min)} & $\approx1.23\cdot 10^3$ & $\approx1.21\cdot 10^3$ & $\approx1.21\cdot 10^3$ & $\approx2.19\cdot 10^3$ & $\approx1.21\cdot 10^3$ & \multicolumn{2}{c|}{$\approx3.2760$} \\ \hline
        \emph{Security Level (Crypto)} & $\star\star\star$ & $\star\star\star$ & $\star\star\star$ & N/A & $\star\star\star$ & \multicolumn{2}{c|}{$\star\star\star$} \\ \hline
        \emph{Health Status Privacy} & $\star$ & $-$ & $-$ & $\star$ & $\star$ & $\star\star$ & $\star\star\star$ \\ \hline
        \emph{Location Privacy (w.r.t Positive)} & $-$ & $-$ & $-$ & $-$ & $-$ & $-$ & $\star\star\star$ \\ \hline
        \emph{Location Privacy (w.r.t Negative)} & $\star\star\star$ & $\star\star\star$ & $\star\star\star$ & $\star\star\star$ & $\star\star\star$ & \multicolumn{2}{c|}{$\star\star\star$} \\ \hline
        \emph{Device Storage Requirements (B)} & $\approx n\cdot140$ & $\approx n\cdot24$ & $\approx n\cdot16$ & $\approx l\cdot10$ & $\approx n\cdot30$ & \multicolumn{2}{c|}{$0$} \\ \hline
        \emph{Crypto Computational Cost (ms)} & $0$ & $\approx24.8973$ & $\approx30.2039$ & $0$ & $0$ & \multicolumn{2}{c|}{$\approx23.3652$} \\ \hline
        \emph{Broadcast TX Overhead (B)} & $f\cdot140$ & $f\cdot24$ & $f\cdot31$ & $0$ & $f\cdot30$ & \multicolumn{2}{c|}{$f\cdot16$} \\ \hline
    \end{tabular}}
\label{tab:comparison}
\end{table*}

\textbf{BlueTrace~\cite{bluetrace}}. BlueTrace is an open-source protocol that is utilized in Singapore's TraceTogether app. It adopts the \ac{BLE} technology, where devices exchange their ephemeral IDs (i.e., beacons) via broadcast and log all encounters in their history logs. When a user is diagnosed as positive, his/her history logs are sent to a central authority, using a secure connection. Even though BlueTrace leverages the decentralized architecture, the ephemeral IDs are generated by the central authority and distributed to the individual devices. As such, the reconciliation function and exposure notification are performed at a centralized location, i.e., BlueTrace is considered a \textit{hybrid} solution. The main cryptographic primitive involved in the computation of the ephemeral IDs is AES-256-GCM.

\textbf{DP-3T~\cite{troncoso2020}}. A large consortium of European researchers, comprising numerous universities and institutions, proposed the Decentralized Privacy-Preserving Proximity Tracing protocol that leverages \ac{BLE} technology to track and log encounters with other users. The contact logs are never transmitted to a central authority, but they are stored only on the client's device. When a user tests positive, his/her ephemeral IDs are transmitted to the central authority. The IDs are generated with symmetric key protocols, such as HMAC-SHA-256 and AES-128-CTR. Finally, the project is completely open-source.

\textbf{Apple/Google~\cite{applegoogle}}. Similar to DP-3T, Apple and Google agreed on a decentralized protocol for contact tracing, based on \ac{BLE} technology. The contact tracing logs do not contain any private information, and ephemeral IDs are only stored on the user's device. From the cryptographic perspective, they adopt HMAC-SHA-256 and AES-128. \textcolor{black}{Note that, Apple/Google is not a complete contact tracing solution; instead, the companies released the exposure notification API as open-source to allow public health authorities to develop their own mobile applications. For example, {\em Immuni}~\cite{immuni} is the Italian State-sponsored official contact tracing app that leverages the Apple/Google framework.}

\textbf{Hamagen~\cite{Hamagen}}. Hamagen was developed by Israel's Ministry of Health to monitor the \textsc{Covid-19} pandemic. It allows the identification of positive patients and people who came in contact with them. Hamagen continuously monitors and logs the user's GPS coordinates on the device (requiring no interaction with other devices). After a user tests positive, and if he/she gives prior consent, their location data is transmitted to the Ministry of Health. All devices periodically download the up-to-date location data and compare them against their own GPS history logs. 
%RDP wording
%Finally, as per the employed adopted tools, the application employs  SHA-256, AES, and HMAC.

%to ensure data integrity they adopt SHA-256. %The application uses symmetric key cryptography such as AES and HMAC.

\textbf{PEPP-PT~\cite{pepppt}}. The Pan-European Privacy-Preserving Proximity Tracing protocol adopts \ac{BLE} to discover and store locally the ephemeral IDs of devices that are in proximity. Similar to BlueTrace, it uses the hybrid architecture by having the health authorities generate the users' beacons. As such, a centralized server collects and processes the contact logs from infected users, and performs the reconciliation process in a centralized manner. The main cryptographic algorithm they employ is AES. This approach also adopts the open-source paradigm. %Finally as confirmed by the authors, ``\textit{code of the PEPP-PT will be under Mozilla Open Source license or similar}``.

{\bf Solutions Comparison.}
Table~\ref{tab:comparison} presents a quantitative comparison of these state-of-the-art protocols for a variety of metrics, such as privacy and operational cost. \textcolor{black}{In our analysis, we consider the health authorities as trusted entities. Otherwise, centralized and hybrid protocols cannot offer any meaningful level of privacy.}
% First, we want to emphasize that a centralized solution, such as PEPP-PT, can theoretically offer perfect privacy guarantees. However, in reality, a central authority that stores detailed contact logs from all users is an obvious target for rogue insiders and malicious hackers. More importantly, these logs may reveal a lot of sensitive information, such as locations visited or health status. As a result, we do not consider centralized solutions as privacy-preserving. \textcolor{black}{Besides, BlueTrace and DP-3T protocols adopt a hybrid architecture where data collection follows the decentralized approach (nothing is disclosed to the authorities)}
\textcolor{black}{In terms of health status privacy, decentralized protocols fail to protect the identity of the infected users, which is a violation of numerous health privacy acts, such as HIPAA and GDPR. Specifically, DP-3T and Apple/Google disclose all the ephemeral IDs that belong to the infected users, which allows an adversary to infer with certainty whether a known ID (i.e., person) has contracted the virus. As for hybrid solutions (BlueTrace and PEPP-PT), they only reveal the user's contact logs and are, thus, more privacy-preserving.} However, the ephemeral ID of the infected individual might be inferred from its absence within a cluster of IDs with the same time/location tags. Hamagen is a GPS-based solution, so it reveals the infected user's entire location history. While the identity of the user may not be immediately clear, background knowledge can be applied to link the published trajectories to a specific individual.

\textcolor{black}{Regarding location privacy, both the decentralized and hybrid protocols offer excellent privacy to the users who never test positive. This is due to the unidirectional flow of information, i.e., the devices only download data from the central authority's server without ever uploading any data of their own. However, a user who tests positive has to disclose some relevant information to the central server.} Usually, the cited disclosure involves publishing ephemeral IDs, contact logs, or GPS coordinates, unfortunately leading, among others, %the end result is 
to a complete compromise of the geographic locations that the user has visited in the near past.

\textcolor{black}{To assess the performance of the discussed solutions in a quantitative manner, we considered the Bluetooth SoC nRF51822 and the GPS SiP nRF9160 (for Hamagen) hardware platforms. We first estimated the energy consumption related to the RF operations (TX and RX), using the platforms' operational specifications, such as voltage and current consumption. For the BLE-based protocols, we assumed a beacon broadcast interval of $500$~ms, and a duty cycle of $50$\% for the scanning function. The energy consumption of each approach is computed as the integral of power over time. For Hamagen, we assumed continuous scanning in low-power mode. As presented in Table~\ref{tab:comparison}, IoTrace is orders of magnitude more efficient than the competing approaches, because it does not need to scan the Bluetooth channel for broadcasted beacons. As per the crypto operations for generating the ephemeral IDs, they are very cheap for all protocols, necessitating $\approx30$~ms to generate the IDs for an entire day (on a Cortex M0 CPU). However, IoTrace is considerably more lightweight, as it does not need to store and actively update a contact list. For the same reason, IoTrace sports the lowest storage requirement.}

\section{Threat Model}
\label{sec:threat}
In this work, we consider a powerful eavesdropping adversary that is capable of collecting all beacons transmitted by the users. The adversary is equipped with a powerful antenna, which can be either a regular Bluetooth handheld device or a \ac{SDR} that is operated through a laptop/smartphone running an SDR-compatible software tool. Additionally, the attacker tags every beacon with a timestamp and the geographic location where it was recorded. As a result, the adversary has a global view of all communications and can pinpoint every beacon to a unique point in space and time, although the beacon cannot be linked to a specific user. We also consider a more involved eavesdropping adversary that is able to get close to a target victim, in order to record beacons that belong to the victim with a very high probability (i.e., there are no other devices in the vicinity, or the adversary uses a directional antenna). Such an adversary is only interested in identifying beacons that are associated with one or more unique individuals.

Finally, we embrace a standard assumption in the literature: the adversary runs in polynomial time and is unable to break the cryptographic protocols (such as symmetric encryption and hashing) that generate the pseudo-random beacons. Based on the aforementioned adversarial model, we consider two types of privacy attacks against the contact tracing system:
\begin{itemize}
    \item \textbf{Location privacy attack:} In this attack, the adversary's objective is to track the movements of one or more users through the collected beacons.
    \item \textbf{Health status privacy attack:} Here, the objective is to correctly infer whether one or more \textit{known} users have contracted the \textsc{Covid-19} virus.
\end{itemize}

\section{Edge Contact Tracing with IoT Devices}
\label{sec:proto}
The novelty of IoTrace lies in the deployment of IoT devices that support the contact tracing tasks at the network's edge, complementing the individual mobile devices. In what follows, Section~\ref{sec:sys_arch} introduces the IoTrace architecture, and Section~\ref{sec:proto_details} describes the contact tracing protocol and the corresponding message flow in the context of a centralized architecture. Section~\ref{sec:scale} discusses an alternative fully distributed approach that also provides a high level of privacy with the use of public-key cryptography.

\subsection{System Architecture}
\label{sec:sys_arch}
The entities involved in the IoTrace architecture are the following:
\begin{itemize}
    \item \emph{User.} A user carrying a smartphone device that runs our contact tracing app. The app simply transmits \ac{BLE} beacons (pseudo-random ephemeral IDs) that are received by the deployed IoT devices. The transmitted beacons are also stored locally on the device for verifying proximity to other users. Unlike existing approaches, the app only operates in transmit-mode, i.e., it does not collect \ac{BLE} beacons from other devices. % In this case, we consider \emph{Alice} as infected user and \emph{Bob} as an entity close to \emph{Alice}.
    \item \emph{Totem.} This is an IoT smart device, equipped with a \ac{BLE} transceiver that collects the beacons transmitted from the users' devices. We also assume that the totem maintains a secure intranet connection to the central authority, where it forwards all the received beacons---in a fashion that could span from batch mode to real-time. In our terminology, we call these beacons \textit{negative}, i.e., they belong to users who have not tested positive. From a practical perspective, a totem could be a simple low-end device, like a Raspberry Pi. 
    \item \emph{Hospital.} This is a medical center that tests users who may possibly have a \textsc{Covid-19} infection. If a user tests positive, the health professionals are permitted to access his/her mobile device and forward the stored beacons to the central authority. We call these beacons \textit{positive}.
    \item \emph{Central authority.} This is a trusted party, whose role is to collect the positive and negative beacons sent by the corresponding hospitals and totems. It is assumed to be always online and ready to provide an updated list of beacons that belong to users who had a close contact with an infected user. In a real scenario, this role can be played by the \emph{Ministry of Health}.
\end{itemize}

\begin{figure}[ht]
    \centering
    \subfigure[Open space: Park.] {
        \centering 
        \includegraphics[width=\linewidth]{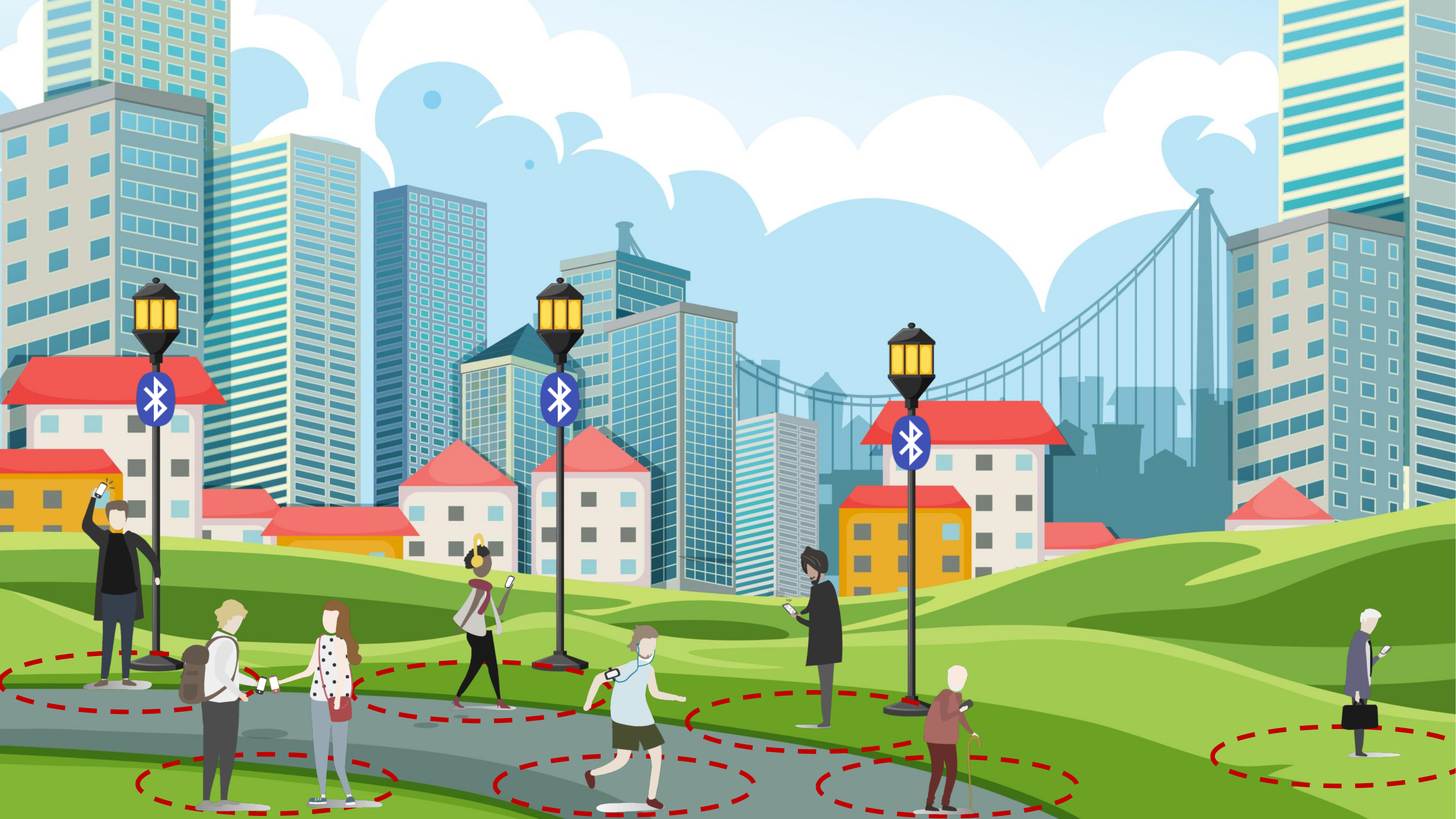}
    }
    \subfigure[Closed space: Shopping mall.] {
        \centering
        \includegraphics[width=\linewidth]{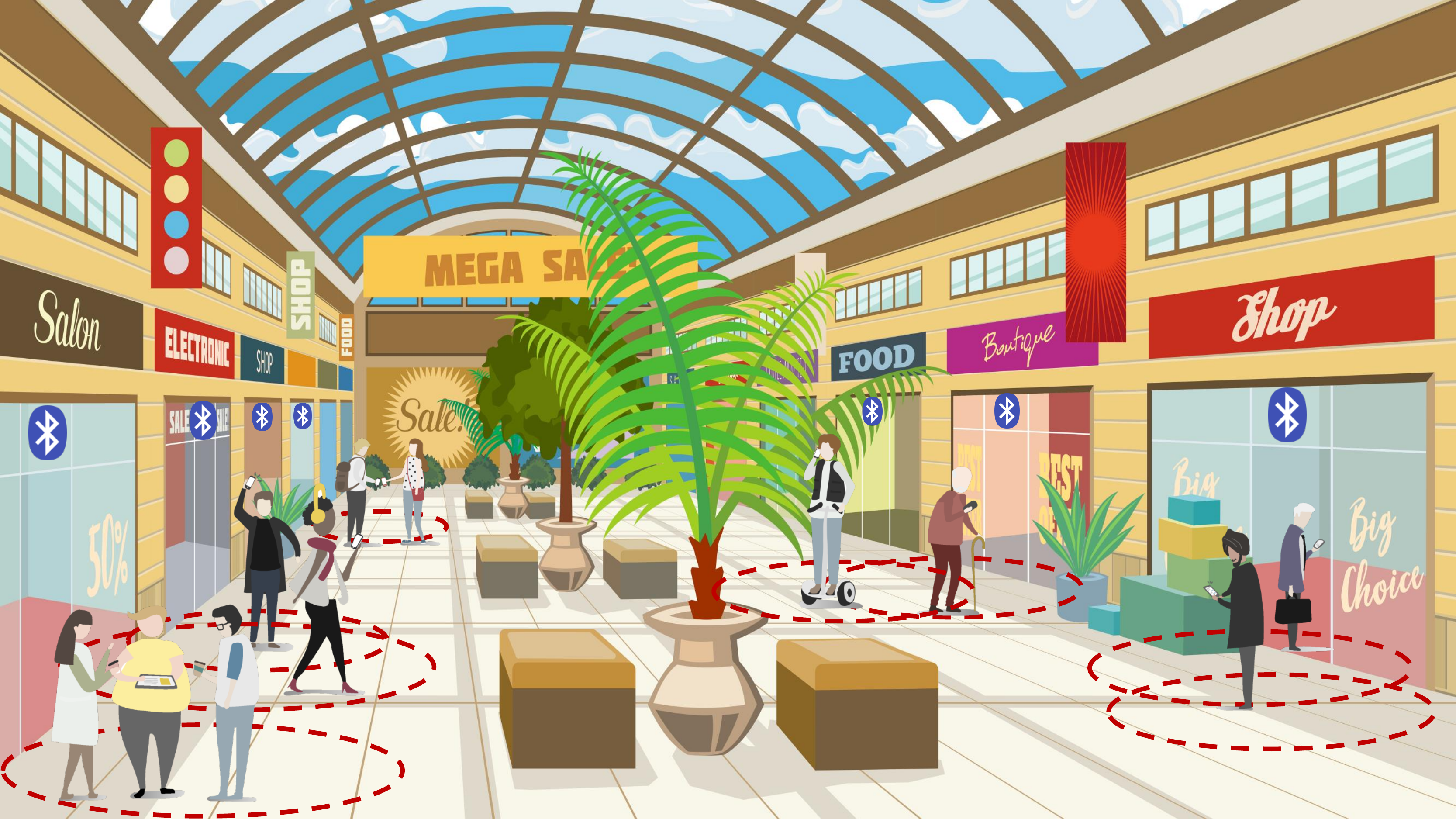}
    }
    \subfigure[Closed space: Office.] {
        \centering 
        \includegraphics[width=\linewidth]{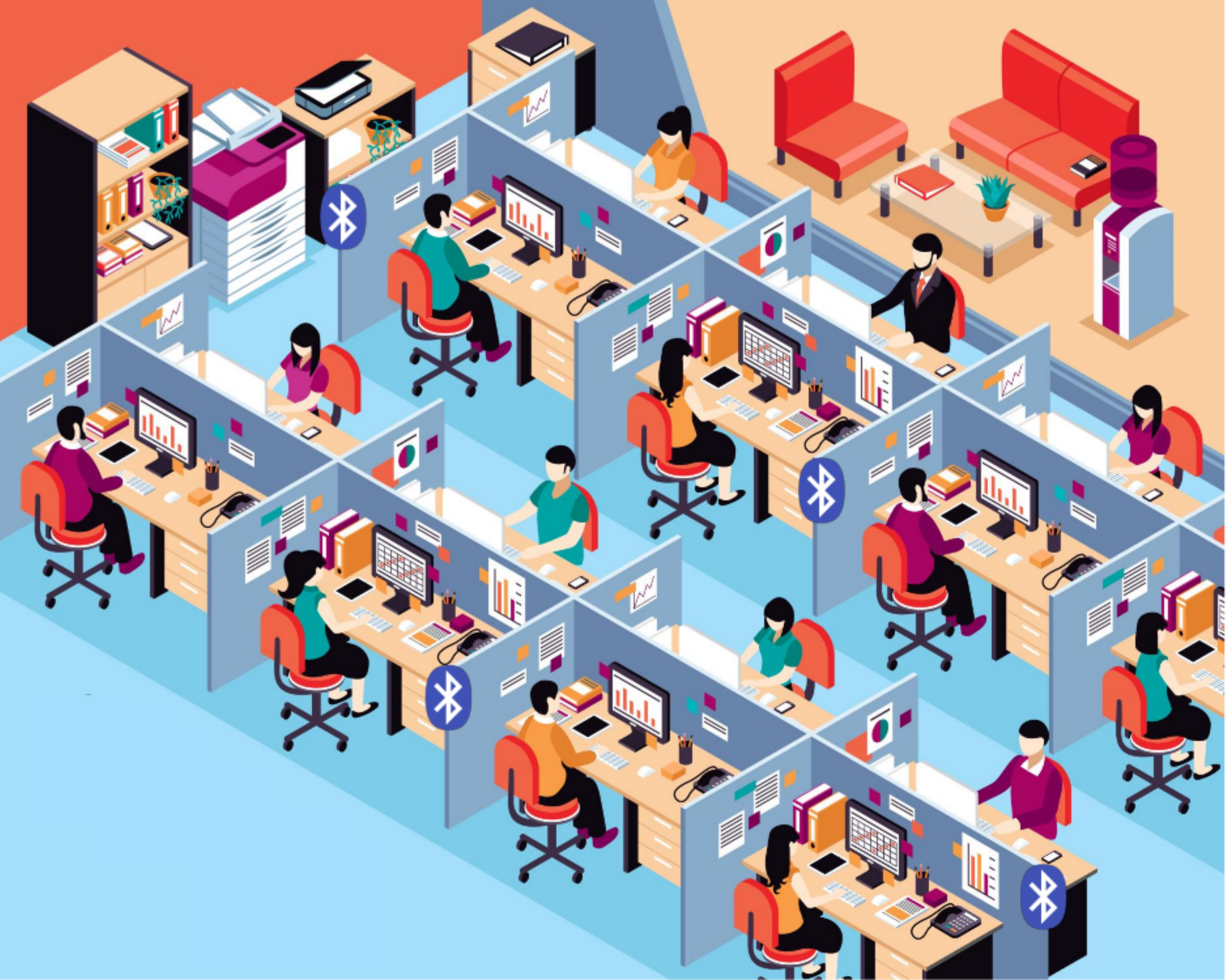}
    }
    \caption{\textcolor{black}{IoTrace infrastructure in different environments: open and closed spaces. A totem is represented by the Bluetooth icon. BLE beacons are transmitted from the smartphones to the IoT totems.}}
    \label{fig:scenarios}
\end{figure}

As shown in Fig.~\ref{fig:scenarios}, the proposed architecture can be adopted in open spaces, like parks, or in closed spaces, like shopping malls and offices.

% \begin{figure}[H]
%   \centering
%   \includegraphics[angle=0, width=\columnwidth]{scenario_park.pdf}
%   \caption{IoTrace in an Open Space: Park.}
%   \label{fig:scenario_park}
% \end{figure}

% \begin{figure}[H]
%   \centering
%   \includegraphics[angle=0, width=\columnwidth]{scenario_mall.pdf}
%   \caption{IoTrace in a Crowded Closed Space: Mall.}
%   \label{fig:scenario_mall}
% \end{figure}

% \begin{figure}[H]
%   \centering
%   \includegraphics[angle=0, width=\columnwidth]{scenario_office.pdf} %, trim=1 1 1 1,clip
%   \caption{IoTrace in a Closed Space: Office.}
%   \label{fig:scenario_office}
% \end{figure}

\begin{figure}[ht]
  \centering
  \includegraphics[angle=0, width=\columnwidth]{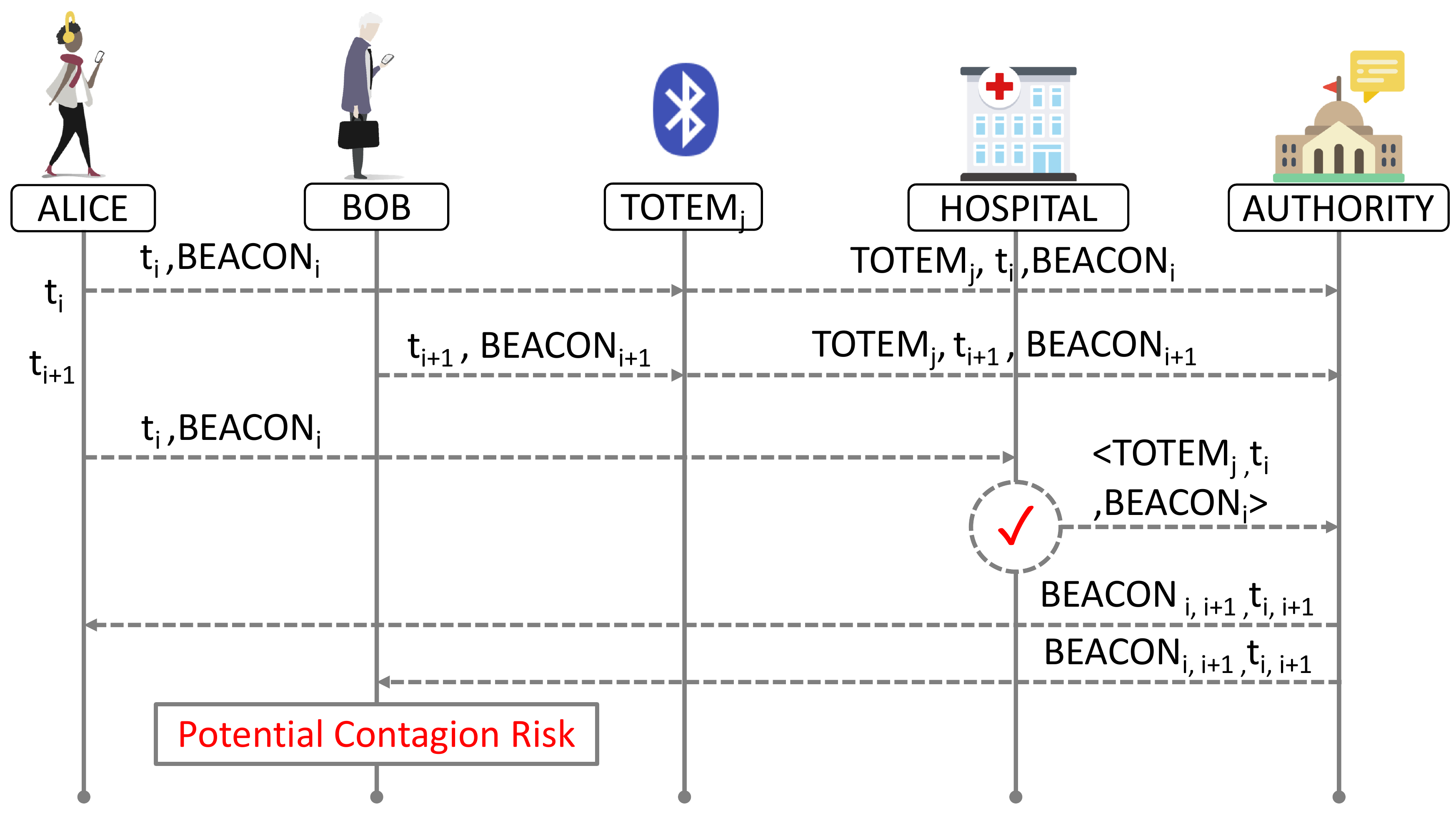} %, trim=1 1 1 1,clip
  \caption{\textcolor{black}{Sequence diagram of the IoTrace protocol. The authority marks Alice as infected by reporting an alert of Potential Contagion Risk to Bob.}}
  \label{fig:protocol}
\end{figure}

\subsection{Protocol Message Flow}
\label{sec:proto_details}
\begin{figure}[ht]
  \centering
  \includegraphics[angle=0, width=\columnwidth]{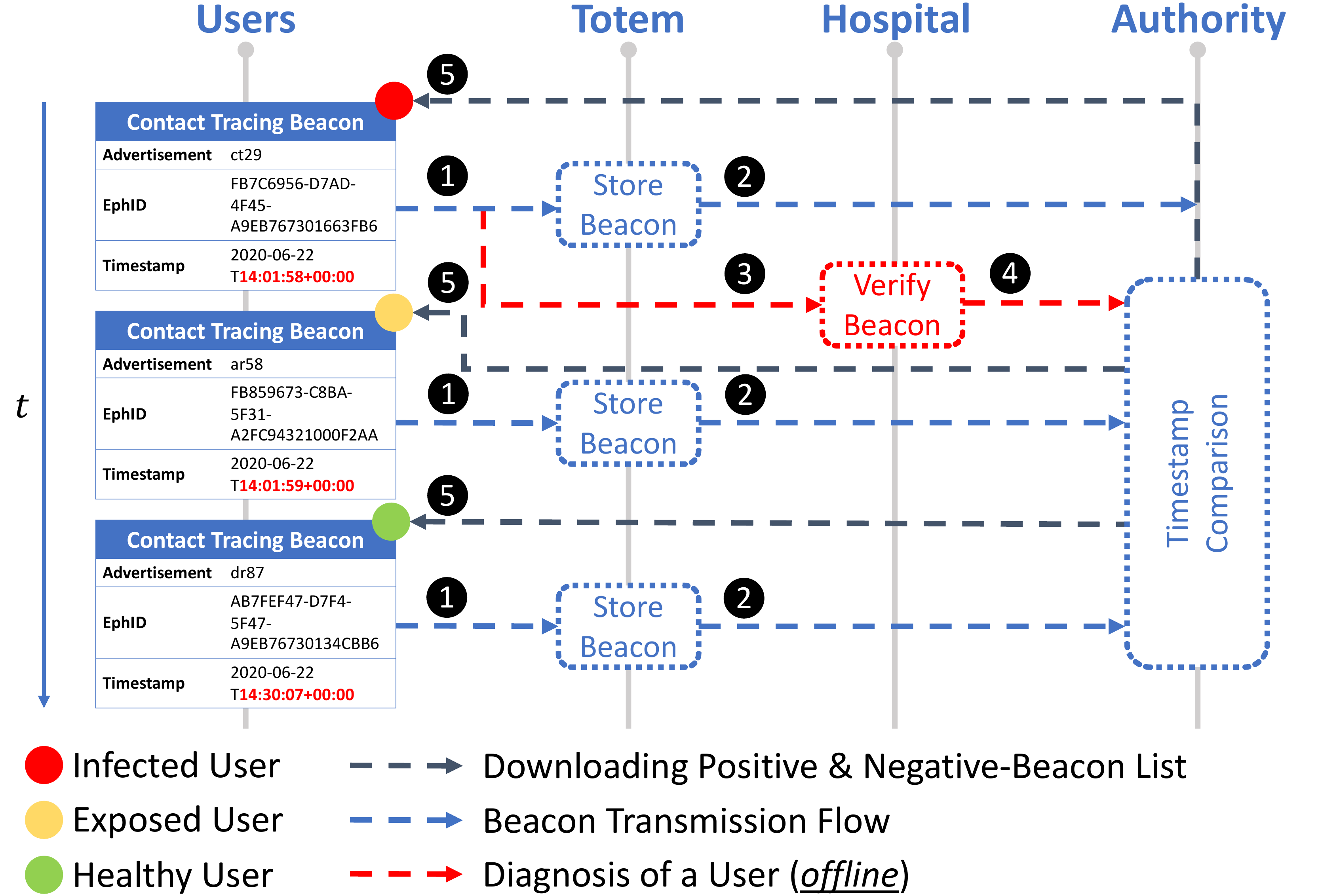}
  \caption{\textcolor{black}{The IoTrace protocol data flow diagram. An exposed user becomes yellow if he/she is close to an infected user based on the BLE beacon's timestamp.}}
  \label{fig:data_flow}
\end{figure}

IoTrace's protocol is illustrated in Fig.~\ref{fig:protocol} and summarized below:
%and consists of the following steps:
\begin{itemize}
    \item Let us assume a generic time-slot $t_i$. At the beginning of the time-slot, the transmitting user (Alice) %/Bob 
    generates a pseudo-random \ac{BLE} beacon, according to some cryptographic primitive, such as AES-128 encryption.
    \item After collecting all beacon(s) within time-slot $t_i$, the totem forwards them to the central authority. The authority stores each beacon as a tuple \textit{<totem-ID, time-slot, beacon>} on its long-term memory. % It is worth noticing that there is no requirement for immediate transmission since it could be more efficient to send a batch of data from multiple time-slots.
    \item Let us assume Alice is diagnosed as positive at an authorized hospital. An authorized health official will access Alice's mobile application to send her recent beacons (e.g., from the last two weeks) to the central authority.
    \item Consider one of Alice's positive beacons that was transmitted at time-slot $\tau$. The central authority will identify all the negative beacons within time-slots $\tau \pm \epsilon$ (at that particular totem), where $\epsilon$ is a time window that depends on the broadcast frequency of the beacons. The list of all negative \textit{and} positive beacons is published online. %and downloaded from the other users.
    \item Finally, Bob downloads from the central authority the list published at the previous step and checks whether his own beacons are in the list. If there is a match, and it is sustained for an amount of time sufficient to declare a potential contagion risk (set by the health authorities), Bob is notified by the app of this possibility.% Note that the check does not necessarily need to be carried out on the mobile, but it could also be carried out on the user's laptop. 
    %an alert is sent to the user with a "High Contagion Risk".
\end{itemize}
The data flow that summarizes the above described operations is depicted in Fig.~\ref{fig:data_flow}.

%\textcolor{black}{The IoTrace architecture is designed by following the \textit{privacy-by-design} methodology introduced with the General Data Protection Regulation (GDPR).}
Compared to previous approaches in the literature, this basic version of IoTrace already provides better protection of the users' health status privacy, since both the positive and the negative beacons are disclosed by the central authority. As a result, IoTrace provides $k$-anonymity~\textcolor{black}{\cite{wang2020}} in terms of health status privacy. That is, if $k$ beacons are published on behalf of a single totem, each beacon has a $1/k$ chance of being the positive one. As per the location privacy guarantees, they are identical to existing decentralized solutions, such as DP-3T and Apple/Google. However, IoTrace has a clear advantage in terms of operational cost for mobile devices.

% Further, we assume that the \ac{BLE} totems must be loosely synchronized. In fact, if time synchronization is not available, the risk of false-positive is higher. For instance, a user is identified as closest to an infected user in the same time window due to the time synchronization misconfiguration.

\subsection{A Privacy-Enhanced Solution}
\label{sec:scale}
We will now show how to significantly enhance the privacy under IoTrace, while leveraging the same architecture. The first improvement is related to the centralized storage of all beacons. To this end, IoTrace can operate in a fully decentralized mode, i.e., the totems will store the received beacons locally, without sending them to the central authority. When a user tests positive for \textsc{Covid-19}, the central authority will forward the positive beacons to all totems and, in turn, the totems will send back to the central authority all negative beacons that fall within the predetermined time window from a positive one. This approach preserves the  privacy guarantees and operational costs for mobile devices while removing the inherent risks of centralized storage.

Our second improvement comes with increased computational and power consumption costs for the mobile devices, but results in a contact tracing solution which is secure against eavesdropping adversaries. The key observation is that the IoT infrastructure is relatively static, so it is easy to store on each mobile device the list of all totem IDs, along with their public-key certificates. Then, instead of transmitting their beacons in cleartext, the mobile devices will first exchange a symmetric key with the totem using the locally stored certificate, and then send their beacons encrypted with that key. The totem will locally decrypt the beacons, and the protocol will continue as described.
%encrypt them with the public-key of the nearest totem. 
Consequently, when the positive/negative beacons are published by the central authority, an eavesdropper cannot link them to a particular totem and time-slot.
%This solution is the first one in the literature to offer very strong privacy guarantees in terms of both location and health status. 
%RDP I am not sure about the sentence above, better to comment (further, we also need to save words)
%Pietro: I agree.
Overall, this latter solution is very flexible, allowing individual users to trade off more privacy with  %in exchange for 
a higher computing cost. % on their smartphones. 
\section{Challenges and Road Ahead}

\label{sec:secpriv_ch}
Contact tracing is, in essence, a surveillance-type application. As such, the security and privacy of the entire system are of paramount importance. In the following sections, we describe the challenges that must be addressed, to make edge contact tracing a secure and privacy-preserving solution.

\subsection{Security Considerations}
\textbf{Edge Security.} In the proposed architecture, an IoT device (totem) represents 
%is configured as 
the edge component between the mobile devices and the hospital/authority. 
Hence, a research direction relevant to our solution, but also of general interest to the IoT domain, arises from the need to reduce the required computations, for instance, adopting lightweight cryptographic protocols
%according to the NIST~\cite{Turan2019} 
to meet the intended security and privacy goals. The most obvious concern with regards to the security of the proposed architecture is the exposure of the totems to physical attacks, due to their being unattended. %Indeed, totems will be widely deployed in areas where attackers can easily access their hardware and read the contents of their memory and non-volatile storage. 
As a result, no sensitive information, such as user beacons or private keys, should be stored in plaintext format. To solve the cited issue, %For instance, 
data at rest could be encrypted with the public key of the central authority. Furthermore, the totem should utilize a secure enclave to perform the necessary cryptographic operations, and all beacons (even when encrypted, as suggested above) should be erased as soon as they are received by the trusted authority. For the case of the fully distributed architecture where the data are stored locally at the totems, additional measures should be implemented to harden their security.
% \textcolor{black}{Another important challenge is the ability of a compromised totem to transmit falsified data to the trusted authority. To this end, the system should incorporate various authentication mechanisms to verify the validity of the submitted beacons. For instance, it would be to design secure two-party protocols (between the trusted authority and a user) that allow users to blindly match their beacons against the server's beacon list---removing the need to publish any type of beacon.}\\

\textbf{Replay and Relay Attacks.} These are active attacks where the adversary eavesdrops on the broadcast beacons and then replays those beacons to many other (even far away) totems. The objective of these attacks is to generate a large number of false contacts such that, if one individual tests positive, the disclosure of his/her beacons will trigger many false positive alerts. Such attacks can be addressed in two different ways. First, the beacon generation protocol may incorporate certain cryptographic protocols to thwart replay attacks. Second, the trusted authority can analyze the collected data and identify fraudulent beacons, e.g., the same beacon appearing in two distant locations in a not time-congruent manner.

% \textcolor{red}{Spiros: I suggest removing DoS and data breaches. They are not applicable only to IoT domain, but are quite general in nature.}

% \textbf{Denial of Service Attacks.} The fixed location of the deployed totems makes them an easy target for DoS attacks. For instance, the adversary may use directional antennas to jam a specific totem or overwhelm it with a large number of random beacons. To this end, we should investigate the applicability of known anti-DoS and anti-jamming techniques to the edge contact tracing use-case. \textcolor{black}{It is worth noticing that a radio jamming attack is always possible even if anti-jamming solutions are adopted. Specifically, a jamming attack towards the \ac{BLE} protocol will blind the smartphones and the totem from the normal transmission and receiving operations.}

% \textbf{Data Breaches.} The trusted authority is a major target for sophisticated hackers that wish to steal the collected beacons. This information is very valuable, as it can be used to disclose sensitive information about the underlying users. For that reason, the centralized server must implement the current state-of-the-art security mechanisms at all levels (network, application, operating system).

\subsection{Privacy Considerations}
\textbf{Linkage and Profiling.} Contact tracing protocols and applications bring with them several privacy concerns, e.g., the misuse of the collected data at the trusted authority, under the centralized and hybrid models. Indeed, a malicious insider with access to all beacons, locations, timestamps, and contact lists, can extract sensitive information about the underlying individuals (such as locations visited, routes, social contacts, etc.). Our proposed architecture makes such attacks less feasible by design, since users do not submit their own contact lists. Instead, all the beacons are aggregated at the distributed totems, which makes it much harder for an adversary to track individuals. Still, an interesting research direction would be to quantify the privacy leakage under the centralized edge contact tracing architecture.

\textbf{Eavesdropping.} Eavesdropping is a passive attack where an adversary with the ability to eavesdrop at a large scale can simply record most beacons that are broadcast by the users. When the list of positive/negative beacons is published, the adversary can identify all the locations that the infected user has visited. This is an attack that none of the existing contact tracing protocols can defend against. To this end, a possible research direction would be to design secure two-party protocols (between the trusted authority and a user) that allow users to blindly match their beacons against the server's beacon list (which will not be published).
Note, however, that our proposal is able to thwart such attacks when the user employs the public-key of the totem to bootstrap a secure channel with the totem itself---as described in Section~\ref{sec:scale}.
%To this end, a possible research direction would be to design secure two-party protocols (between the trusted authority and a user) that allow users to blindly match their beacons against the server's beacon list (which will not be published). 

% \textbf{Path Disclosure.} 
% %The main difference between our architecture and the existing decentralized contact tracing solutions, is that contacts are recorded through an intermediate stepping stone (the totem) instead of a peer-to-peer manner. 
% Totems are located at fixed geographic locations, so a beacon that is stored in a series of totems will disclose this partial path to an adversary. Therefore, to preserve path privacy, beacons should change very frequently (to an extreme, every beacon should be transmitted exactly one time), or beacons should be encrypted, as in Section \ref{sec:scale}.

\subsection{Technology Considerations}
\textbf{Localization Accuracy.} Most technologies adopted for contact tracing rely on the \ac{RSSI}. With the help of a radio-propagation model, this feature is useful in estimating the distance between the transmitter and the receiver nodes. Unfortunately, several factors can affect the accuracy of distance estimation, including radio noise, obstacles, multipath reflection and shadowing effects, or environmental factors like rain, temperature, and humidity. Therefore, Bluetooth \ac{RSSI} may produce a large number of false positives and false negatives. To this end, alternative features like Angle-of-Arrival, Time Difference of Arrival, and Time of Arrival should be investigated. Furthermore, thanks to the vast amount of available data, AI  %machine/deep learning 
algorithms could be employed on the edge devices to improve the localization accuracy of the Bluetooth technology.

\textbf{Communication Technologies.}
% Although \ac{BLE} has been the choice of election by both the scientific and the industrial community, despite the widely known security and privacy issues of wireless technologies applied in this context, 
While \ac{BLE} is the de facto choice for all contact tracing solutions in the literature, 
we believe that more research is needed on different communication technologies. In particular, Ultra-wideband (UWB) carriers as well as acoustic channels and ultrasonic sound waves could be employed to improve the accuracy, privacy, and reliability of proximity tracing~\cite{Caprolu2020}.

% \textbf{Device Maintenance.} Our proposed architecture necessitates the widescale deployment of cheap IoT devices. However, the effectiveness of contact tracing depends on the continuous availability of these devices. To this end, the limited energy storage of IoT devices (just a battery) may increase the maintenance cost significantly, as they may require frequent battery replacements. % Contact tracing solutions require that people have a charged smartphone and to always carry it. If the smartphone's battery is low, or people do not have a smartphone with them, they cannot rely on the current solutions provided in the literature.
% As such, it would be interesting to pursue the development of proximity tracing tools that rely on battery-less IoT devices.

% \textbf{Data Standardization and Interoperability.} Data standardization format is another major concern among the contact tracing solutions. Indeed, a standard template to represent the data would allow 
% interoperability, or at least an easier tracking task, stemming  %easy tracking among 
% from the information collected by different applications--- hence, even between States, possibly easing traveling.

\subsection{Social Considerations\label{sec:social_cons}}
\textcolor{black}{
\textbf{Accessibility.} IoTrace shifts a significant portion of the energetic and computational costs of contact tracing to the IoT edge devices and/or the centralized server. As a result, the corresponding mobile application can be easily deployed on low-cost devices that would otherwise be unable to participate in the contact tracing network. This will increase considerably the accessibility of the solution to the general public.}
% The technology adopted in our solution relies on \ac{BLE}. A user without a smart device like a bracelet, smartphone, smartwatch, or any other device equipped with \ac{BLE} and enabling the contact tracing paradigm cannot be used with our solution. To this end, alternative \ac{BLE}-based low-cost devices can be designed to provide more accessibility for a large number of people, even those who do not have a smartphone. Another social user concern could be the lack of data accessibility collected by the totem which can be adopted for research or public health services.}

\textcolor{black}{
\textbf{Usability.} The usability of existing solutions is primarily hindered by the shortened battery life that users experience. 
%On less powerful devices, the computational cost may also become a factor, especially during the expensive reconciliation process, where the contact list is compared against the newly published ``infected'' beacons (which could be in the order of millions). 
We have shown that energy consumption under IoTrace is reduced by multiple orders of magnitude. Additionally, the reconciliation process is mostly performed at the health authorities and/or IoT devices. As such, we argue that IoTrace's mobile app would be extremely lightweight, and therefore, would not affect the user's experience---hence increasing the chance of adoption.}
% The UX interface of the contact tracing app provided with our solution should be simple and modern to guarantee a good user experience without affecting the usability of its device. Furthermore, the application should be not invasive and annoying users with advertisements and notifications. As a research challenge, it would be interesting to adopt our solution without providing a specific application for the contact tracing. Further, the usability of a contact tracing solution could drain the smartphones' batteries. IoTrace guarantees low energy consumption by stretching the smartphone's battery life and improving usability.}

\textcolor{black}{
\textbf{Trust.} In addition to usability, trust (or the lack of) is the deciding factor that discourages people from actively using existing contact tracing apps. To this end, IoTrace's superior privacy guarantees could motivate more users to install and actively use the app. Furthermore, by releasing the app's code as open-source, we can further ease the public's concern with respect to privacy and security.}

\textcolor{black}{\subsection{Limitations\label{sec:limitations}} The major limitation of IoTrace is the cost to deploy, operate, and maintain the IoT infrastructure. Indeed, the IoT devices must be connected to a fixed power supply and have access to a cellular/cable network infrastructure, in order to communicate with the health authorities. As such, we envision that a practical implementation would employ cheap, Raspberry Pi like devices, that would cost somewhere between \$$10$--\$$20$ each. For $100,000$ devices, the cost would rise to a couple of million dollars, which is very reasonable for a large city. We should emphasize that IoTrace would only be deployed in crowded areas, such as shopping malls, public transportation venues, airports, stadiums, parks, etc. Additionally, the government may offer incentives to individual business owners to install and maintain their own IoT devices, thus expanding the range of IoTrace's network.}

\textcolor{black}{Despite the cited costs, IoTrace has the following advantages that make it a very attractive solution for contact tracing: (i) significant energy savings for the mobile devices, as they can operate in transmit-only mode; (ii) superior privacy guarantees; (iii) better proximity tracing accuracy stemming from a moderately dense deployment of IoT sensors (improved localization with techniques like triangulation and trilateration); (iv) reduced computational and storage requirements for the mobile devices, allowing the app to work seamlessly on cheap devices; (v) flexibility on behalf of the health authorities, because IoTrace does not enforce any constraint on the distance (or duration) that qualifies a digital encounter as a legitimate contact.}

\section{Conclusion}
\label{sec:conclusion}
%Contact tracing protocols are promising solutions to identify people who may have come into contact with an infected person. 

% In this paper, we have provided several contributions: we highlighted the main features %affecting some 
% of the most relevant % important 
% solutions for  contact tracing, % systems, 
% \textcolor{black}{by presenting the}  relationship with computational requirements, security, and privacy levels. In particular, we have reported a clear unbalance when it comes to the privacy guarantees provided to the different categories of users. To this end,
In this paper, we proposed IoTrace, a novel IoT-based architecture for contact tracing, that addresses some of the most important limitations of existing solutions: it provides a balance between the level of privacy for the different user categories; it reduces the overhead on the end-user device in terms of energy consumption and computational cost; it enhances location privacy; and, it is scalable and flexible---allowing to accommodate different contact tracing models, from the purely decentralized to the centralized one. We believe that the novelty of the proposal, as well as its striking properties and flexibility, has the potential to pave the way for further research.
%start a novel research thread.
%resummarized the most important security and privacy challenges, directions, and countermeasures, to be addressed by academia and industry towards the development of secure contact tracing protocols.

\section*{Acknowledgments}
%The authors would like to thank the anonymous reviewers for their comments and suggestions, which helped to improve the quality of the manuscript.
The authors would like to thank the anonymous reviewers, that helped improving the quality of the paper.
This publication was partially supported by awards NPRP 11S-0109-180242 
from the QNRF-Qatar National Research Fund, a member of The Qatar Foundation. The information and views set out in this publication are those of the authors and do not necessarily reflect the official opinion of the QNRF.

%\balance
%RDP refs with 'Accessed' should be updated to Nov; ref. [7] is incomplete

\bibliographystyle{IEEEtran}
\bibliography{references}

\begin{thebibliography}{10}
\providecommand{\url}[1]{#1}
\csname url@samestyle\endcsname
\providecommand{\newblock}{\relax}
\providecommand{\bibinfo}[2]{#2}
\providecommand{\BIBentrySTDinterwordspacing}{\spaceskip=0pt\relax}
\providecommand{\BIBentryALTinterwordstretchfactor}{4}
\providecommand{\BIBentryALTinterwordspacing}{\spaceskip=\fontdimen2\font plus
\BIBentryALTinterwordstretchfactor\fontdimen3\font minus
  \fontdimen4\font\relax}
\providecommand{\BIBforeignlanguage}[2]{{%
\expandafter\ifx\csname l@#1\endcsname\relax
\typeout{** WARNING: IEEEtran.bst: No hyphenation pattern has been}%
\typeout{** loaded for the language `#1'. Using the pattern for}%
\typeout{** the default language instead.}%
\else
\language=\csname l@#1\endcsname
\fi
#2}}
\providecommand{\BIBdecl}{\relax}
\BIBdecl

\bibitem{jhu20}
{\textcolor{black}{John Hopkins University}}, ``{\textcolor{black}{Coronavirus
  Resource Center}},'' \textcolor{black}{\url{https://coronavirus.jhu.edu/}},
  \textcolor{black}{November} \textcolor{black}{2020},
  \textcolor{black}{(Accessed: 2021-1-1)}.

\bibitem{Ting2020}
D.~{Shu Wei Ting} \emph{et~al.}, ``{Digital technology and {COVID}-19},''
  \emph{{Nature Medicine}}, vol.~26, no.~4, pp. 459--461, Mar. 2020.

\bibitem{Cecilia2020}
J.~M. {Cecilia} \emph{et~al.}, ``\textcolor{black}{{Mobile crowdsensing
  approaches to address the COVID-19 pandemic in Spain}},''
  \emph{\textcolor{black}{{IET Smart Cities}}}, vol. \textcolor{black}{2}, no.
  \textcolor{black}{2}, pp. \textcolor{black}{58--63}, \textcolor{black}{2020}.

\bibitem{Garg2020}
L.~{Garg} \emph{et~al.}, ``{Anonymity Preserving IoT-Based COVID-19 and Other
  Infectious Disease Contact Tracing Model},'' \emph{{IEEE Access}}, vol.~8,
  pp. 159\,402--159\,414, 2020.

\bibitem{Hamagen}
\BIBentryALTinterwordspacing
{Israeli Health Ministry}. (2020) {Hamagen}. (Accessed: 2021-1-1). [Online].
  Available:
  \url{https://govextra.gov.il/ministry-of-health/hamagen-app/download-en/}
\BIBentrySTDinterwordspacing

\bibitem{Tang20}
Q.~Tang, ``{Privacy-Preserving Contact Tracing: current solutions and open
  questions},'' \emph{{IACR} Cryptol. ePrint Arch.}, vol. 2020, p. 426, 2020.

\bibitem{baumgrtner2020mind}
L.~Baumg{\"a}rtner \emph{et~al.}, ``{Mind the GAP: Security \& Privacy Risks of
  Contact Tracing Apps},'' \emph{arXiv e-prints}, Jun. 2020.

\bibitem{Lu2019}
Y.~Lu \emph{et~al.}, ``{Internet of Things (IoT) Cybersecurity Research: A
  Review of Current Research Topics},'' \emph{IEEE Internet of Things Journal},
  vol.~6, no.~2, pp. 2103--2115, 2019.

\bibitem{bluetrace}
J.~{Bay} \emph{et~al.}, ``{BlueTrace: A privacy-preserving protocol for
  community-driven contact tracing across borders},'' \emph{Government
  Technology Agency-Singapore, Tech. Rep}, 2020.

\bibitem{troncoso2020}
``{Decentralized Privacy-Preserving Proximity Tracing: Overview of Data
  Protection and Security},''
  \url{https://github.com/DP-3T/documents/blob/master/DP3T\%20White\%20Paper.pdf},
  2020, (Accessed: 2021-1-1).

\bibitem{applegoogle}
\BIBentryALTinterwordspacing
{Apple Google}. (2020) {Privacy-Preserving Contact Tracing}. (Accessed:
  2021-1-1). [Online]. Available:
  \url{https://www.apple.com/covid19/contacttracing}
\BIBentrySTDinterwordspacing

\bibitem{pepppt}
\BIBentryALTinterwordspacing
{PEPP-PT Team}. (2020) {Pan-European Privacy-Preserving Proximity Tracing}.
  (Accessed: 2021-1-1). [Online]. Available: \url{https://www.pepp-pt.org/}
\BIBentrySTDinterwordspacing

\bibitem{immuni}
{\textcolor{black}{Italian Ministry of Health}},
  ``{\textcolor{black}{Immuni}},''
  \textcolor{black}{\url{https://www.immuni.italia.it/}},
  \textcolor{black}{June} \textcolor{black}{2020}, \textcolor{black}{(Accessed:
  2021-1-1)}.

\bibitem{wang2020}
J.~{Wang} \emph{et~al.}, ``{\textcolor{black}{Achieving Personalized
  $k$-Anonymity-Based Content Privacy for Autonomous Vehicles in CPS}},''
  \emph{{\textcolor{black}{IEEE Transactions on Industrial Informatics}}}, vol.
  \textcolor{black}{16}, no. \textcolor{black}{6}, pp.
  \textcolor{black}{4242--4251}, \textcolor{black}{2020}.

\bibitem{Caprolu2020}
M.~{Caprolu} \emph{et~al.}, ``{Short-Range Audio Channels Security: Survey of
  Mechanisms, Applications, and Research Challenges},'' \emph{IEEE
  Communications Surveys Tutorials}, pp. 1--1, 2020.

\end{thebibliography}

\section*{Biographies}

\begin{IEEEbiography}[{\includegraphics[width=1in,height=1.25in,clip,keepaspectratio]{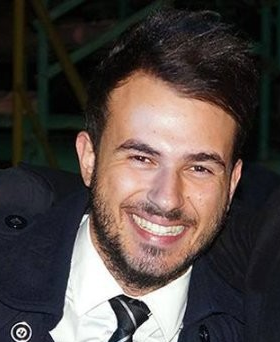}}]{Pietro Tedeschi}
is PhD Student at HBKU-CSE. He received his Master's degree with honors in Computer Engineering at Politecnico di Bari, Italy. He worked as Security Researcher at CNIT, Italy, for the EU H2020 SymbIoTe. His research interests cover security issues in UAVs, Wireless, IoT, and Cyber-Physical Systems. 
\end{IEEEbiography}

\begin{IEEEbiography}[{\includegraphics[width=1in,height=1.25in,clip,keepaspectratio]{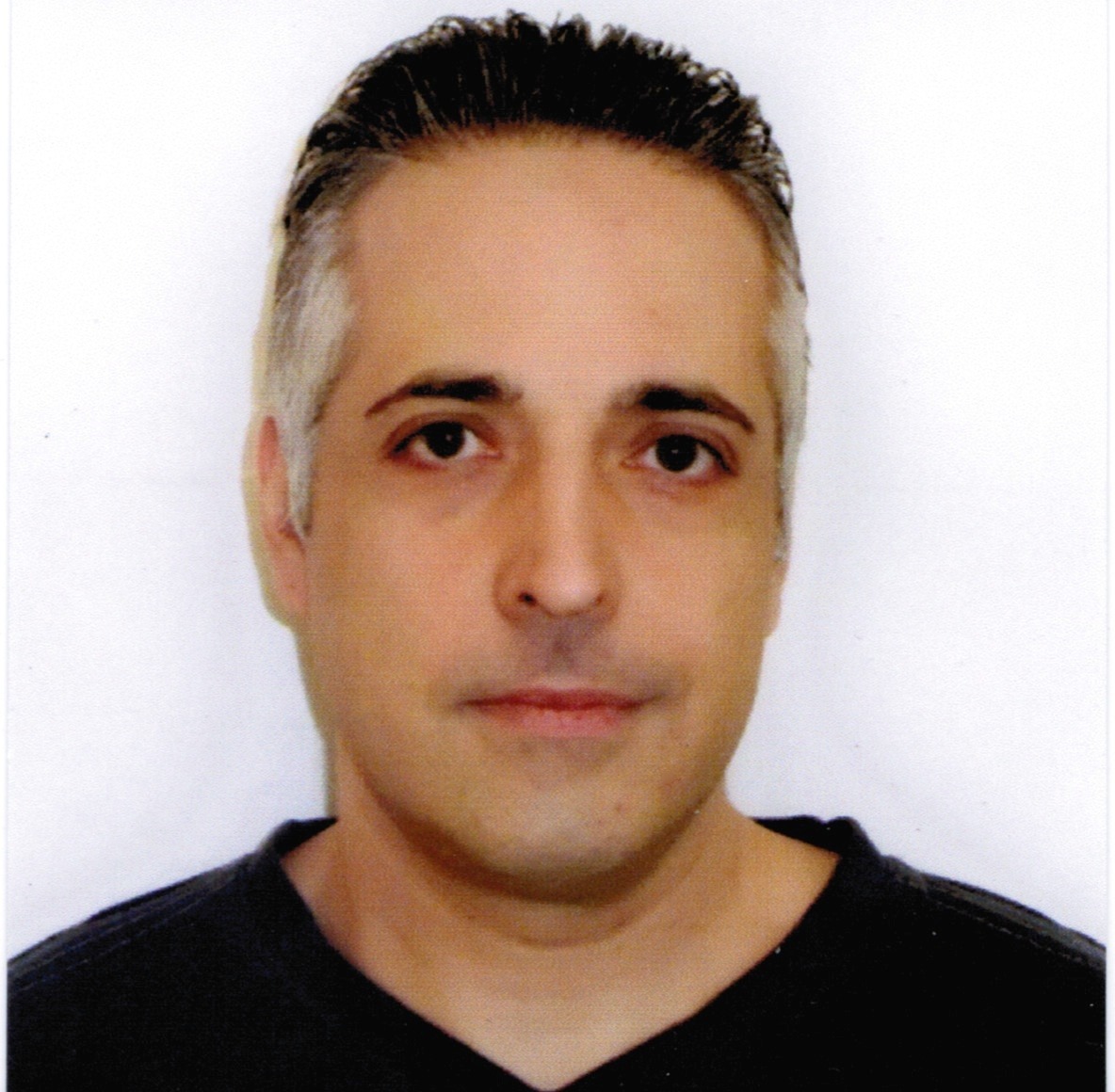}}]{Spiridon Bakiras}
is associate professor of cybersecurity at HBKU-CSE. His research interests include Security and Privacy, Applied Cryptography, and Spatiotemporal Databases. He held teaching and research positions at Michigan Technological University, the City University of New York, the University of Hong Kong, and the Hong Kong University of Science and Technology. He is a recipient of the U.S. National Science Foundation CAREER award. 
\end{IEEEbiography}

\begin{IEEEbiography}[{\includegraphics[width=1in,height=1.25in,clip,keepaspectratio]{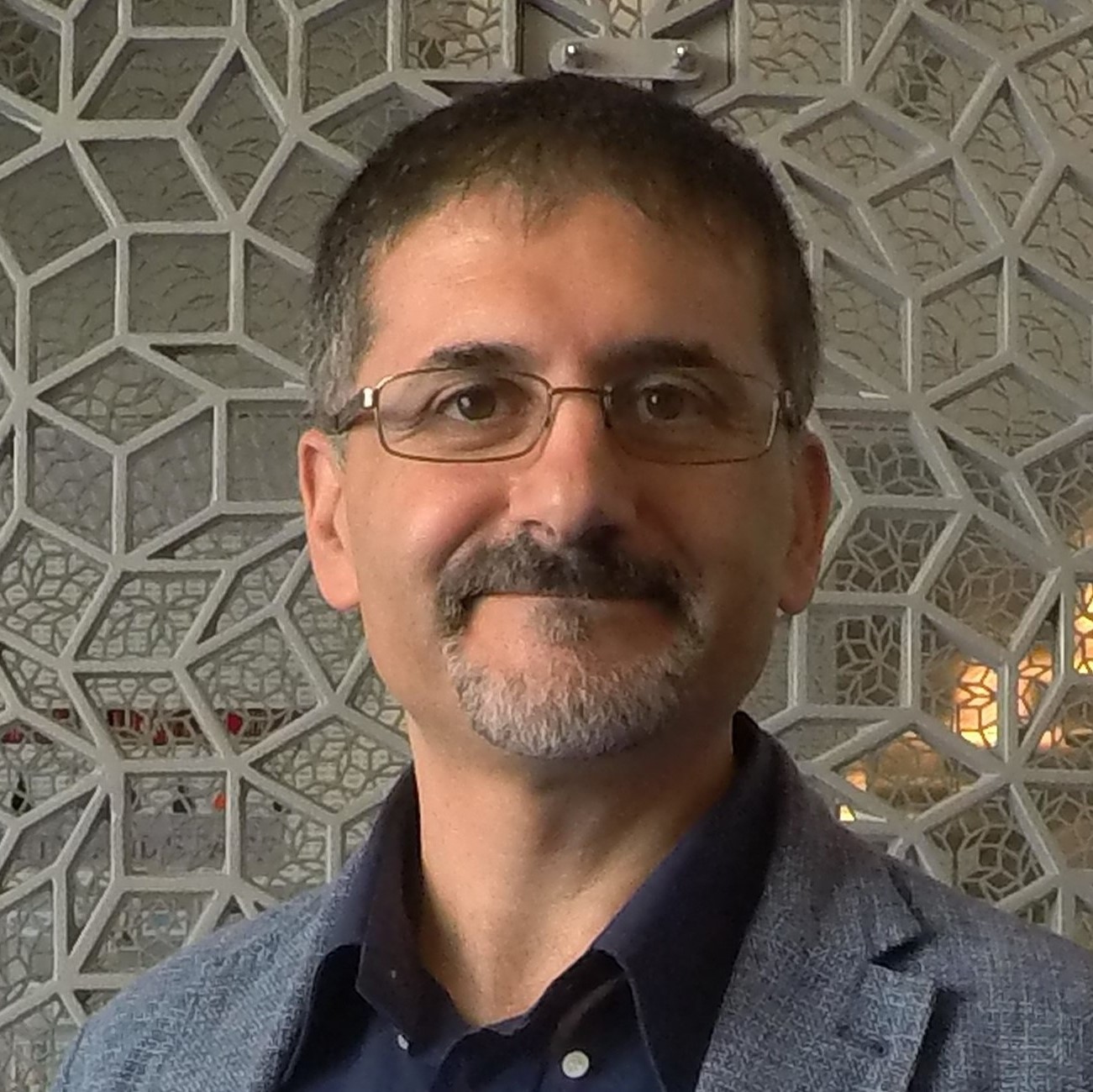}}]{Roberto Di Pietro,} ACM Distinguished Scientist, is full professor of Cybersecurity at HBKU-CSE. His research interests include Distributed Systems Security, Wireless Security, OSN Security, and Intrusion Detection.
%, leading to $220+$ scientific publications and patents. 
 In 2011-2012 he was awarded a Chair of Excellence from University Carlos III, Madrid. In 2020 he received the Jean-Claude Laprie Award for having significantly influenced the theory and practice of Dependable Computing. %As for Google Scholar, he has been totaling 9100$+$ citations, with \mbox{h-index=46}, and \mbox{i-index}=128.
\end{IEEEbiography}

\end{document}